# An artificially intelligent magnetic resonance spectroscopy quantification method: Comparison between QNet and LCModel on the cloud computing platform CloudBrain-MRS

Meijin Lin [a,b,1], Lin Guo [c,1], Dicheng Chen [b,d], Jianshu Chen [c], Zhangren Tu [b,d], Xu Huang [e], Jianhua Wang [e], Ji Qi [f], Yuan Long [f], Zhiguo Huang [f], Di Guo [g], Xiaobo Qu [e,d,b,c,*], Haiwei Han [e,*]

[a] *College of Ocean & Earth Sciences, Xiamen University, Xiamen 361102, China.*

[b] *Fujian Provincial Key Laboratory of Plasma and Magnetic Resonance, Xiamen University, Xiamen 361102, China.*

[c] *National Integrated Circuit Industry Education Integration Innovation Platform, Xiamen University, Xiamen 361100, China.*

[d] *Department of Electronic Science, Xiamen University, Xiamen 361102, China.*

[e] *Department of Radiology, the First Affiliated Hospital of Xiamen University, School of Medicine, Xiamen University, Xiamen 361100, China.*

[f] *China Mobile (Suzhou) Software Technology Company Limited, Suzhou 215163, China.*

[g] *School of Computer and Information Engineering, Fujian Engineering Research Center for Medical Data Mining and Application, Xiamen University of Technology, Xiamen 361024, China.*

---

\* Corresponding author.
 E-mail addresses: hanminghui360@163.com (H. Han), quxiaobo@xmu.edu.cn (X. Qu)
[1] Shared first authorship.

# Abstract

*Objctives:* This work aimed to statistically compare the metabolite quantification of human brain magnetic resonance spectroscopy (MRS) between the deep learning method QNet and the classical method LCModel through an easy-to-use intelligent cloud computing platform CloudBrain-MRS.

*Materials and Methods:* In this retrospective study, two 3 T MRI scanners Philips Ingenia and Achieva collected 61 and 46 *in vivo* $^1$H magnetic resonance (MR) spectra of healthy participants, respectively, from the brain region of pregenual anterior cingulate cortex from September to October 2021. The analyses of Bland-Altman, Pearson correlation and reasonability were performed to assess the degree of agreement, linear correlation and reasonability between the two quantification methods.

*Results:* Fifteen healthy volunteers (12 females and 3 males, age range: 21-35 years, mean age ± standard deviation = 27.4 ± 3.9 years) were recruited. The analyses of Bland-Altman, Pearson correlation and reasonability showed high to good consistency and very strong to moderate correlation between the two methods for quantification of total N-acetylaspartate (tNAA), total choline (tCho), and inositol (Ins) (relative half interval of limits of agreement = 3.04%, 9.3%, and 18.5%, respectively; Pearson correlation coefficient $r$ = 0.775, 0.927, and 0.469, respectively). In addition, quantification results of QNet are more likely to be closer to the previous reported average values than those of LCModel.

*Conclusion:* There were high or good degrees of consistency between the quantification results of QNet and LCModel for tNAA, tCho, and Ins, and QNet generally has more reasonable quantification than LCModel.

**Keywords**

Magnetic resonance spectroscopy; Brain metabolite quantification; Consistency analysis; QNet; LCModel; Cloud computing platform CloudBrain-MRS

# 1. Introduction

Magnetic resonance (MR) spectroscopy (MRS) is widely used to measure the concentration of brain metabolites [1], and changes in concentration reflect neurological disorders [2-4]. Reliable MRS quantification [1,5,6] will be greatly helpful for correct clinical diagnosis. The widely used classic method LCModel analyzes *in vivo* spectrum as a Linear Combination of Model spectra of metabolite solutions *in vitro*, which improves the fitting accuracy [7-11].

Recently, deep learning has been used in the MR field for fast signal reconstruction, denoising, and metabolite quantification [12-21]. However, one of its bottlenecks is the difficulty acquiring enough training data due to lengthy and expensive measurement, or even worse, the ground-truth labels, such as metabolite concentrations of *in vivo*, are unavailable. An effective solution to this problem is synthetic data. Huge amount of physics-driven synthetic data can be produced without cost through physical evolution or analytical models to train neural network. Deep learning MR methods have achieved good performance of high quality and reliability with synthetic data [18,21-23]. Synthetic data learning has been evidenced powerful for MRS in denoising, baseline removal, and quantification, etc [24-26]. Among these deep learning methods which resort to synthetic data to train network, for instance, QNet [21] is a state-of-the-art method that incorporates the conventional least square concentration estimation into deep learning scheme to greatly improve the model generalization, even though, application of QNet to clinical data requires careful investigation.

To use a quantification method independent from a platform, no matter LCModel or QNet, users need to have specialized knowledge of MRS, know how to code, install the program and its dependencies, and configure the program environment, etc., which greatly increases the difficulty to use them [9,27]. What is more, the subsequent analysis, such as statistical analysis and plotting figures, usually involves many steps that are cumbersome and time-consuming, if all steps are done manually. These inconveniences could be handled well with a cloud computing platform [28-30]. As an intelligent cloud computing platform for MRS, CloudBrain-MRS [28] integrates both QNet and LCModel to provide both quantification and statistical results after simple setup, saving all the cumbersome steps with a highly efficient workflow.

In this work, through CloudBrain-MRS, we assess the consistency, and compare the reasonability, and assess the correlation of the MRS quantification results between QNet and LCModel with MRS data of healthy subjects.

## 2. Materials and methods

*2.1. Participant overview*

The protocol of this retrospective study was approved by the Hospital Institutional Review Board of the First Affiliated Hospital of Xiamen University, and written informed consent was obtained from all the volunteers prior to the experiments. Fifteen healthy volunteers (12 females and 3 males, age range: 21-35 years, mean age $\pm$ standard deviation (SD) = 27.4 $\pm$ 3.9 years) were recruited. All the volunteers were physically healthy and had no mental illnesses and self-reported no history of head trauma, substance abuse or dependence, or neurological disorders. Two 3 T MRI scanners, Philips Ingenia and Achieva, were used to scan each subject repeatedly on different days to collect *in vivo* MRS spectra from the brain region of pregenual anterior cingulate cortex (pgACC) at the First Affiliated Hospital of Xiamen University from September to October 2021. The spectra of poor visual quality are excluded. Finally, 61 and 46 *in vivo* MRS spectra from Philips Ingenia and Philips Achieva, respectively, were used in this study.

*2.2. MRS data acquisition*

A Philips Ingenia and a Philips Achieva 3T MR whole-body scanner with an 8-channel head coil were used to acquire single voxel data. The fast spin-echo sequence was adopted to acquire T1-weighted brain images for positioning the pgACC voxel, and the single voxel PRESS pulse sequence "svs_press" was used to acquire MR spectra with an excitation module for water suppression. The pulse sequence parameters are as follows: TR = 2000 ms; TE = 35 ms; bandwidth = 2000 Hz; number of points = 1024; number of averages = 128; VOI = 20 mm×20 mm×20 mm.

*2.3. Data processing on the platform CloudBrain-MRS*

After the MRS data were gathered from the MRI scanners, the convenient platform CloudBrain-MRS was utilized to perform the quantification of the metabolite concentrations with both QNet and LCModel, and together with the reasonability analysis of both methods and the consistency analysis between the two methods in the meanwhile. This can be done just by simple setup and choosing "consistency analysis" on the panel of CloudBrain-MRS (Fig. 1). The fitting spectra of both methods can be showed on CloudBrain-MRS to check whether their fittings are good with small residual (Fig. 2).

*2.4. Statistical analysis of consistency, Pearson correlation and reasonability*

To assess the degree of agreement between the two MRS quantification methods QNet and LCModel in quantifying brain metabolite concentrations, the Bland-Altman analysis [31] was integrated in CloudBrain-MRS to show the differences and 95% limits of agreement (LoA).

Specifically, CloudBrain-MRS first calculates the difference between each pair of values from two methods of a sample, and the normality of the differences is examined by Shapiro-Wilk (SW) test. If the differences are normally distributed, that is, the *p*-value of the SW test is larger than 0.05, then the SD of the differences can be used to define the limits of agreement (LoA), and Bland-Altman analysis then can applied to reliably assess the degree of agreement between the two methods [31].

The mean of the differences, i.e. bias, reflects the systematic difference between the two methods. 95% LoA is calculated as the mean $\pm 1.96 \times$ SD of the differences [31], thus, half interval of LoA equals 1.96 SD of the differences. The smaller the better for both bias and the interval of LoA. In this study, the average of the two methods of a sample, denoted as average, is used as a reference to assess the levels of both the systematic difference and the interval of LoA of the two methods. That is, relative bias (bias/average) and relative half interval of LoA (1.96 SD of differences/average) are used to represent the systematic bias and the difference interval, respectively.

Along with Bland-Altman analysis, Pearson correlation coefficient is used to measure the linear association between the concentrations estimated by the two methods.

Box-plot of the estimated concentrations is used to show the reasonability of the quantification result with the metabolite level ranges reported from the literature [32].

We focus on the analysis of the relative concentrations of the following five major metabolites: (1) total N-acetylaspartate (tNAA) = N-acetyl aspartate + N-acetyl aspartyl glutamate, (2) total choline (tCho) = free choline + glycerophosphorylcholine + phosphorylcholine, (3) Glx = glutamate + glutamine, (4) inositol (Ins), aka myo-inositol (mI), and (5) glutathione (GSH). For relative concentration, total creatine (tCr), i.e. creatine + phosphocreatine, is used as a reference.

## 3. Results

*3.1. Analyses of Bland-Altman and Pearson correlation*

In this study, the consistency between QNet and LCModel for the quantification of the relative concentrations of five major metabolites (tNAA/tCr, tCho/tCr, Glx/tCr, Ins/tCr, and GSH/tCr) in the human brain was assessed through the analyses of Bland-Altman (Fig. 3-4 and Tables 1-3). Meanwhile, the linear association was assessed by the Pearson correlation coefficient (Tables 1-3).

For the data gathered from Philips Ingenia (Fig. 3 and Table 1), the relative half intervals of LoA are quite small for tCho (7.45%), and small for tNAA (12.64%) and Ins (19.19%), which demonstrates that there are high or relatively high degrees of agreement between QNet and LCModel in quantification of tNAA, Ins, and tCho. Likewise, the Pearson correlation coefficients ($r$) statistically ($p < 0.01$) demonstrate the linear correlation between the two methods are strong for tCho (0.893) and for tNAA (0.767) and moderate for Ins (0.502).

Table 1 also shows that, with the mean concentration of the both methods of the sample as the reference, the relative biases are quite small for tNAA (1.74%) and Ins (3.74%), and small for tCho (10.03%), which means the systematic differences between QNet and LCModel are small for the three metabolites.

For the data gathered from Philips Achieva (Fig. 4 and Table 2), the results are similar. The relative half intervals of LoA are quite small for tCho (4.76%), and small for tNAA (9.26%) and Ins (15.91%), and the Pearson correlation coefficients ($r$) demonstrate the linear correlation between the two methods are strong for tCho ($r = 0.943$) and for tNAA ($r = 0.811$) and moderate for Ins ($r = 0.418$) all with $p < 0.01$.

Similarly, Table 2 also shows that, the relative biases are quite small for tNAA (4.02%), and small for Ins (8.63%) and tCho (9.34%), which means the systematic differences between QNet and LCModel are small for the three metabolites.

When the data from the two scanners are put together to perform the analysis, and in order to obtain the normal distribution of the differences, 8 spectra from Philips Ingenia were discarded, that is, only 53 spectra from Philips Ingenia and all the 46 spectra from Philips Achieva were used. The results are similar as well (Table 3).

*3.2. Reasonability analysis*

Boxplots along with the previous reported metabolite level ranges were used to assess the reasonability of metabolite concentrations quantified by QNet and LCModel with all the spectra acquired by both the scanners (Fig. 5). Figure 5 shows that, according to the normal ranges of metabolite concentrations reported in the literature [32], QNet is more reasonable in general. Specifically, for tCho, Glx and GSH, the boxes of the QNet are closer to the mean values than those of LCModel; and for tNAA and Ins, the two methods are almost the same. What is more, for QNet, the concentrations of all metabolites except GSH are within the normal ranges, while for LCModel, a small part of points of tCho and Glx, besides most points of GSH, are outside the ranges.

## 4. Discussion

In this study, the SW normality test of the differences between the two methods, QNet and LCModel, was performed to ensure the Bland-Altman analysis is valid to assess the degree of agreement of the two methods in quantifying the metabolites in the human brain. The smaller the range of LoA, the better the agreement is. However, Bland-Altman analysis only define the intervals of agreement, it does not say whether those limits are acceptable or not [33]. Acceptable limits must be defined a priori, based on clinical necessity, biological considerations or other goals [33]. In this study, there is no pre-defined value for the maximum acceptable difference. Meanwhile, the true values cannot be obtained in this study, and the mean of the two measurements are the best estimate we have. Therefore, in this study, the average metabolite concentration of the two methods of all the points in a sample was used as a reference to assess how large the systematic difference and the interval of LoA are.

Bland-Altman analysis shows that, the systematic difference, in term of the ratio of bias (i.e. the average differences) to the average concentration of the two methods are very small (up to 5%) for tNAA, and small (up to 10%) for Ins and tCho, but large (about 40%) for GSH and the largest (up to about 60%) for Glx. However, even if there is a consistent systematic bias, it is a simple matter to adjust by subtracting the mean difference from the measurements.

After the systematic difference is removed, the two methods can be exchangeable if the interval of differences is small. In terms of the ratio of (1.96 SD of differences/average concentration of the two methods), Bland-Altman analysis shows that, the relative half intervals of LoA are quite small for tNAA and tCho usually with the level of 10%, and relatively small for Ins with the level of 20%, but larger for Glx with the level of 30%, and the largest for GSH with the level of 50%.

The Pearson correlation coefficients also show the similar results with Bland-Altman analysis, that is, the quantification values of tCho, tNAA and Ins are highly or moderately correlated between the two methods, but those of Glx and GSH are not. The low degrees of agreement for Glx and GSH maybe due to that the chemical shifts of the two metabolites are too close to separate from each other at 3 T.

The reasonability analysis of the quantification shows that QNet seems usually more reasonable than LCModel, since the values measured by QNet are more likely to be within the normal ranges or closer to the mean values. The boxplots show that the tNAA values are very similar between the two methods, and as well the values of tCho and Ins are similar.

There are limitations in this study. First, it is very difficult to detect the true values of the brain metabolite levels of *in vivo*, since the metabolite signals are very weak, and what is more, many of the MRS peaks overlap with each other at 3 T. Therefore, it is hard to tell which method is more accurate if there are differences between methods. Second, the sample size may not be sufficient to fully represent all possible measurement scenarios, and there is no patient data to fully evaluate the two methods. Future work may increase the sample size and diversify the subjects.

## 5. Conclusion

This study used the cloud computing platform CloudBrain-MRS to assess the consistency and compare the reasonability of the two methods QNet and LCModel for MRS quantification of five major metabolite (tNAA, tCho, Glx, Ins and GSH) concentrations relative to tCr. All the analyses of Bland-Altman, Pearson correlation and reasonability demonstrate that the two methods have small differences or highly to moderately correlate for the quantification of tNAA, tCho, and Ins, which means the two methods may be used interchangeable for the quantification of these three metabolites after removal of the systematic differences. In addition, QNet are usually more reasonable than LCModel according to the reported normal metabolite ranges. Therefore, the deep learning method QNet may be a good alternative MRS quantification method. In addition, the highly efficient platform CloudBrain-MRS streamlines the usage of the quantification methods QNet or LCModel, which can further facilitate the clinical application of MRS.

**Funding**

This work was supported in part by the National Natural Science Foundation of China (grants 62331021, 62122064 and 62371410), the Natural Science Foundation of Fujian Province of China (grant 2023J02005), Industry-University Cooperation Projects of the Ministry of Education of China (grant 231107173160805), National Key R&D Program of China (grant 2023YFF0714200), the President Fund of Xiamen University (grant 20720220063), and Nanqiang Outstanding Talent Program of Xiamen University.

**References**

[1] J.B. Poullet, D.M. Sima, S. Van Huffel, MRS signal quantitation: A review of time and frequency-domain methods, J. Magn. Reson. 195 (2008) 134-144. http://doi.org/10.1016/j.jmr.2008.09.005.

[2] J.S. Axford, F.A. Howe, C. Heron, J.R. Griffiths, Sensitivity of quantitative $^1$H magnetic resonance spectroscopy of the brain in detecting early neuronal damage in systemic lupus erythematosus, Ann. Rheum. Dis. 60 (2001) 106-111. http://doi.org/10.1136/ard.60.2.106.

[3]     F.A. Howe, S.J. Barton, S.A. Cudlip, M. Stubbs, D.E. Saunders, M. Murphy, P. Wilkins, K.S. Opstad, V.L. Doyle, M.A. McLean, B.A. Bell, J.R. Griffiths, Metabolic profiles of human brain tumors using quantitative in vivo $^1$H magnetic resonance spectroscopy, Magn. Reson. Med. 49 (2003) 223-232. http://doi.org/10.1002/mrm.10367.

[4]     A.Q. Lin, J.X. Shou, X.Y. Li, L. Ma, X.H. Zhu, Metabolic changes in acute cerebral infarction: Findings from proton magnetic resonance spectroscopic imaging, Exp. Ther. Med. 7 (2013) 451-455. http://doi.org/10.3892/etm.2013.1418.

[5]     L.G. Kaiser, M. Marjańska, G.B. Matson, I. Iltis, S.D. Bush, B.J. Soher, S. Mueller, K. Young, $^1$H MRS detection of glycine residue of reduced glutathione in vivo, J. Magn. Reson. 202 (2010) 259-266. http://doi.org/10.1016/j.jmr.2009.11.013.

[6]     X.-H. Zhu, M. Lu, W. Chen, Quantitative imaging of brain energy metabolisms and neuroenergetics using in vivo X-nuclear $^2$H, $^{17}$O and $^{31}$P MRS at ultra-high field, J. Magn. Reson. 292 (2018) 155-170. http://doi.org/10.1016/j.jmr.2018.05.005.

[7]     S.W. Provencher, Estimation of metabolite concentrations from localized in vivo proton NMR spectra, Magn. Reson. Med. 30 (1993) 672-679. http://doi.org/10.1002/mrm.1910300604.

[8]     J.W. van der Veen, R. de Beer, P.R. Luyten, D. van Ormondt, Accurate quantification of in vivo $^{31}$P NMR signals using the variable projection method and prior knowledge, Magn. Reson. Med. 6 (1988) 92-98. http://doi.org/10.1002/mrm.1910060111.

[9]     S.W. Provencher, Automatic quantitation of localized in vivo $^1$H spectra with LCModel, NMR Biomed. 14 (2001) 260-264. http://doi.org/10.1002/nbm.698.

[10]    G. Tackley, Y. Kong, R. Minne, S. Messina, A. Winkler, A. Cavey, R. Everett, G.C. DeLuca, A. Weir, M. Craner, I. Tracey, J. Palace, C.J. Stagg, U. Emir, An in-vivo $^1$H-MRS short-echo time technique at 7T: Quantification of metabolites in chronic multiple sclerosis and neuromyelitis optica brain lesions and normal appearing brain tissue, NeuroImage 238 (2021) 118225. http://doi.org/10.1016/j.neuroimage.2021.118225.

[11]    G. Rodríguez-Nieto, O. Levin, L. Hermans, A. Weerasekera, A.C. Sava, A. Haghebaert, A. Huybrechts, K. Cuypers, D. Mantini, U. Himmelreich, S.P. Swinnen, Organization of neurochemical interactions in young and older brains as revealed with a network approach: Evidence from proton magnetic resonance spectroscopy ($^1$H-MRS), NeuroImage 266 (2023) 119830. http://doi.org/10.1016/j.neuroimage.2022.119830.

[12]    H.H. Lee, H. Kim, Intact metabolite spectrum mining by deep learning in proton magnetic resonance spectroscopy of the brain, Magn. Reson. Med. 82 (2019) 33-48. http://doi.org/10.1002/mrm.27727.

[13]    J. Songeon, S. Courvoisier, L. Xin, T. Agius, O. Dabrowski, A. Longchamp, F. Lazeyras, A. Klauser, In vivo magnetic resonance $^{31}$P-spectral analysis with neural networks: $^{31}$P-SPAWNN, Magn. Reson. Med. 89 (2022) 40-53. http://doi.org/10.1002/mrm.29446.

[14]    Z. Iqbal, D. Nguyen, M.A. Thomas, S. Jiang, Deep learning can accelerate and quantify simulated localized correlated spectroscopy, Sci. Rep. 11 (2021) 8727. http://doi.org/10.1038/s41598-021-88158-y.

[15]    A. Shamaei, J. Starcukova, Z. Starcuk, Physics-informed deep learning approach to quantification of human brain metabolites from magnetic resonance spectroscopy data, Comput. Biol. Med. 158 (2023) 106837. http://doi.org/10.1016/j.compbiomed.2023.106837.

[16]    Z. Wang, D. Guo, Z. Tu, Y. Huang, Y. Zhou, J. Wang, L. Feng, D. Lin, Y. You, T. Agback, V. Orekhov, X. Qu, A sparse model-inspired deep thresholding network for exponential signal

reconstruction-application in fast biological spectroscopy, IEEE Trans. Neural. Netw. Learn. Syst. 34 (2023) 7578-7592. http://doi.org/10.1109/tnnls.2022.3144580.

[17] Y. Huang, J. Zhao, Z. Wang, V. Orekhov, D. Guo, X. Qu, Exponential signal reconstruction with deep hankel matrix factorization, IEEE Trans. Neural. Netw. Learn. Syst. 34 (2021) 6214-6226. http://doi.org/10.1109/tnnls.2021.3134717.

[18] Q. Yang, Z. Wang, K. Guo, C. Cai, X. Qu, Physics-driven synthetic data learning for biomedical magnetic resonance: The imaging physics-based data synthesis paradigm for artificial intelligence, IEEE Signal Proc. Mag. 40 (2023) 129-140. http://doi.org/10.1109/msp.2022.3183809.

[19] D. Chen, Z. Wang, D. Guo, V. Orekhov, X. Qu, Review and prospect: Deep learning in nuclear magnetic resonance spectroscopy, Chem. - Eur. J. 26 (2020) 10391-10401. http://doi.org/10.1002/chem.202000246.

[20] D. Guo, X. Chen, M. Lu, W. He, S. Luo, Y. Lin, Y. Huang, L. Xiao, X. Qu, Review and prospect: Applications of exponential signals with machine learning in nuclear magnetic resonance, Spectroscopy 38 (2023) 22-32. http://doi.org/org/10.56530/spectroscopy.yx1073b8.

[21] D. Chen, M. Lin, H. Liu, J. Li, Y. Zhou, T. Kang, L. Lin, Z. Wu, J. Wang, J. Li, J. Lin, X. Chen, D. Guo, X. Qu, Magnetic resonance spectroscopy quantification aided by deep estimations of imperfection factors and macromolecular signal, IEEE Trans. Biomed. Eng. 71 (2024) 1841-1852. http://doi.org/10.1109/tbme.2024.3354123.

[22] X. Qu, Y. Huang, H. Lu, T. Qiu, D. Guo, T. Agback, V. Orekhov, Z. Chen, Accelerated nuclear magnetic resonance spectroscopy with deep learning, Angew. Chem. Int. Ed. Engl. 59 (2020) 10297-10300. http://doi.org/10.1002/anie.201908162.

[23] D. Chen, W. Hu, H. Liu, Y. Zhou, T. Qiu, Y. Huang, Z. Wang, M. Lin, L. Lin, Z. Wu, J. Wang, H. Chen, X. Chen, G. Yan, D. Guo, J. Lin, X. Qu, Magnetic resonance spectroscopy deep learning denoising using few in vivo data, IEEE Trans. Comput. Imaging. 9 (2023) 448-458. http://doi.org/10.1109/tci.2023.3267623.

[24] M. Dziadosz, R. Rizzo, S.P. Kyathanahally, R. Kreis, Denoising single MR spectra by deep learning: miracle or mirage?, Magn. Reson. Med. 90 (2023) 1749-1761. http://doi.org/10.1002/mrm.29762.

[25] F. Lam, X. Peng, Z.-P. Liang, High-dimensional MR spatiospectral imaging by integrating physics-based modeling and data-driven machine learning: Current progress and future directions, IEEE Signal Proc. Mag. 40 (2023) 101-115. http://doi.org/10.1109/msp.2022.3203867.

[26] Y. Li, Z. Wang, R. Sun, F. Lam, Separation of metabolites and macromolecules for short-TE $^1$H-MRSI using learned component-specific representations, IEEE Trans. Med. Imaging. 40 (2021) 1157-1167. http://doi.org/10.1109/tmi.2020.3048933.

[27] Vanhamme, A. Van den Boogaart, S. Van Huffel, Improved method for accurate and efficient quantification of MRS data with use of prior knowledge, J. Magn. Reson. 129 (1997) 35-43. http://doi.org/10.1006/jmre.1997.1244.

[28] X. Chen, J. Li, D. Chen, Y. Zhou, Z. Tu, M. Lin, T. Kang, J. Lin, T. Gong, L. Zhu, J. Zhou, O.-Y. Lin, J. Guo, J. Dong, D. Guo, X. Qu, CloudBrain-MRS: An intelligent cloud computing platform for in vivo magnetic resonance spectroscopy preprocessing, quantification, and analysis, J. Magn. Reson. 358 (2023) 107601. http://doi.org/10.1016/j.jmr.2023.107601.

[29] Y. Zhou, Y. Wu, Y. Su, J. Li, J. Cai, Y. You, J. Zhou, D. Guo, X. Qu, Cloud-magnetic resonance

imaging system: In the era of 6G and artificial intelligence, Magn. Reson. Lett. (2024) 200138. http://doi.org/10.1016/j.mrl.2024.200138.

[30] D. Guo, S. Li, J. Liu, Z. Tu, T. Qiu, J. Xu, L. Feng, D. Lin, Q. Hong, M. Lin, Y. Lin, X. Qu, CloudBrain-NMR: An intelligent cloud-computing platform for NMR spectroscopy processing, reconstruction, and analysis, IEEE Trans. Instrum. Meas. 73 (2024) 1-11. http://doi.org/10.1109/tim.2024.3415787.

[31] J.M. Bland, D.G. Altman, Statistical methods for assessing agreement between two methods of clinical measurement, Lancet 327 (1986) 307-310. http://doi.org/10.1016/S0140-6736(86)90837-8.

[32] V. Govindaraju, K. Young, A.A. Maudsley, Proton NMR chemical shifts and coupling constants for brain metabolites, NMR Biomed. 13 (2000) 129-153. http://doi.org/10.1002/1099-1492(200005)13:3%3C129::aid-nbm619%3E3.0.co;2-v.

[33] D. Giavarina, Understanding Bland Altman analysis, Biochem. Med. 25 (2015) 141-151. http://doi.org/10.11613/bm.2015.015.

**Table 1**

The Bland-Altman analysis and Pearson correlation coefficients *r* of the relative concentrations from the two methods applied to the data from Philips Ingenia.

| Metabolite | tNAA/tCr | tCho/tCr | Glx/tCr | Ins/tCr | GSH/tCr |
|---|---|---|---|---|---|
| Bias | 0.020 | -0.028 | -1.124 | 0.033 | -0.096 |
| Average | 1.139 | 0.282 | 1.805 | 0.891 | 0.288 |
| Bias/average (%) | 1.74 | -10.03 | -62.30 | 3.74 | -33.44 |
| Upper LoA | 0.164 | -0.007 | 0.691 | 0.204 | 0.023 |
| Lower LoA | -0.124 | -0.050 | -1.558 | -0.138 | -0.216 |
| 1.96 SD | 0.144 | 0.021 | 0.433 | 0.171 | 0.120 |
| 1.96 SD/average (%) | 12.64 | 7.45 | 24.00 | 19.19 | 41.67 |
| *r* | 0.767** | 0.893** | 0.246 | 0.502** | 0.340** |

Note: tNAA = total N-acetylaspartate, tCho = total choline, Glx = glutamate + glutamine, Ins = inositol, GSH = glutathione, tCr = total creatine, LoA = limits of agreement. Bias is the average of the differences of the two methods. Average is the average concentration of the two methods of all subjects in the sample. SD is the standard deviation of the differences of the two method, and thus 1.96 SD is the half interval of LoA. *r* is Pearson correlation coefficient. ** is the correlation with $p < 0.01$.

**Table 2**

The Bland-Altman analysis and Pearson correlation coefficients $r$ of the relative concentrations from the two methods applied to the data from Philips Achieva.

| Metabolite | tNAA/tCr | tCho/tCr | Glx/tCr | Ins/tCr | GSH/tCr |
|---|---|---|---|---|---|
| Bias | 0.048 | -0.025 | -0.760 | 0.073 | -0.114 |
| Average | 1.194 | 0.268 | 1.696 | 0.846 | 0.281 |
| Bias/average (%) | 4.02 | -9.34 | -44.81 | 8.63 | -40.57 |
| Upper LoA | 0.159 | -0.012 | -0.383 | 0.208 | 0.021 |
| Lower LoA | -0.062 | -0.037 | -1.136 | -0.061 | -0.249 |
| 1.96 SD | 0.111 | 0.013 | 0.377 | 0.135 | 0.135 |
| 1.96 SD/average (%) | 9.26 | 4.67 | 22.20 | 15.91 | 48.04 |
| $r$ | 0.811** | 0.946** | 0.379** | 0.418** | 0.329* |

Note: tNAA = total N-acetylaspartate, tCho = total choline, Glx = glutamate + glutamine, Ins = inositol, GSH = glutathione, tCr = total creatine, LoA = limits of agreement. Bias is the average of the differences of the two methods. Average is the average concentration of the two methods of all subjects in the sample. SD is the standard deviation of the differences of the two methods, and thus 1.96 SD is the half interval of LoA. $r$ is Pearson correlation coefficient. ** is the correlation with $p < 0.01$. * is the correlation with $p < 0.05$.

**Table 3**

The Bland-Altman analysis and Pearson correlation coefficients *r* of the relative concentrations from the two methods applied to the data from both scanners.

| Metabolite | tNAA/tCr | tCho/tCr | Glx/tCr | Ins/tCr | GSH/tCr |
| --- | --- | --- | --- | --- | --- |
| Bias | 0.035 | -0.025 | -0.948 | 0.073 | -0.102 |
| Average | 1.168 | 0.274 | 1.752 | 0.871 | 0.283 |
| Bias/average (%) | 2.998 | -9.124 | -54.125 | 8.386 | -36.042 |
| Upper LoA | 0.170 | 0.010 | -0.405 | 0.216 | 0.024 |
| Lower LoA | 0.099 | -0.041 | -1.491 | -0.106 | -0.228 |
| 1.96 SD | 0.036 | 0.026 | 0.055 | 0.055 | 0.126 |
| 1.96 SD/average (%) | 3.041 | 9.307 | 31.002 | 18.495 | 44.523 |
| *r* | 0.775** | 0.927** | 0.090 | 0.469** | 0.322** |

Note: tNAA = total N-acetylaspartate, tCho = total choline, Glx = glutamate + glutamine, Ins = inositol, GSH = glutathione, tCr = total creatine, LoA = limits of agreement. Bias is the average of the differences of the two methods. Average is the average concentration of the two methods of all subjects in the sample. SD is the standard deviation of the differences of the two methods, and thus 1.96 SD is the half interval of LoA. *r* is Pearson correlation coefficient. ** is the correlation with $p < 0.01$.

**Fig. 1.** Setup on the panel of CloudBrain-MRS at http://csrc.xmu.edu.cn/CloudBrain.html. This figure is a screenshot from CloudBrain-MRS

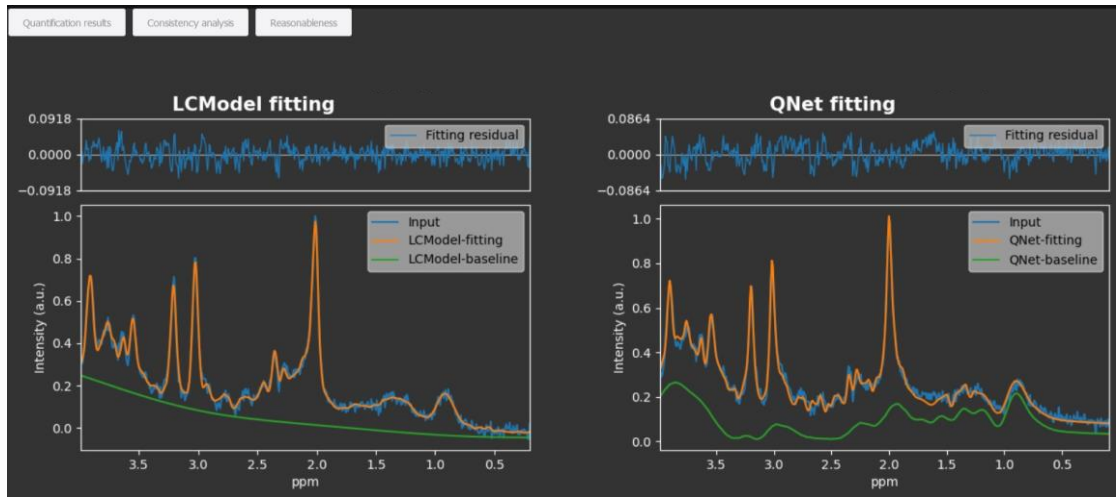

**Fig. 2.** The fitting spectra and their residuals of LCModel (left) and QNet (right) from a subject spectrum. This figure is a screenshot from CloudBrain-MRS

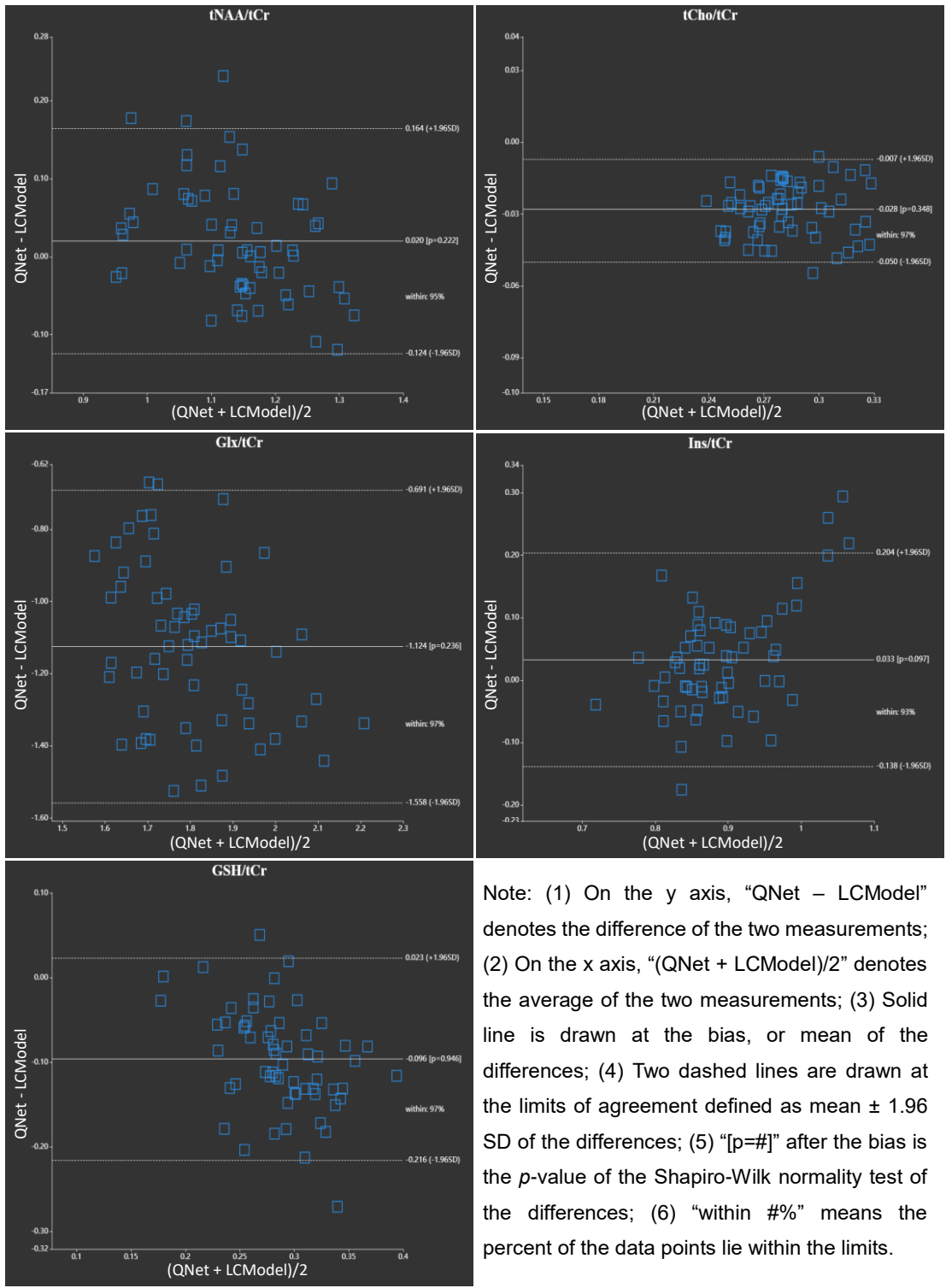

Note: (1) On the y axis, "QNet – LCModel" denotes the difference of the two measurements; (2) On the x axis, "(QNet + LCModel)/2" denotes the average of the two measurements; (3) Solid line is drawn at the bias, or mean of the differences; (4) Two dashed lines are drawn at the limits of agreement defined as mean ± 1.96 SD of the differences; (5) "[p=#]" after the bias is the *p*-value of the Shapiro-Wilk normality test of the differences; (6) "within #%" means the percent of the data points lie within the limits.

**Fig. 3.** Bland-Altman analysis for the 61 spectra acquired by scanner Philips Ingenia between QNet and LCModel for quantification of the five metabolites relative to tCr of tNAA, tCho, Glx, Ins, and GSH. A box indicates a spectrum in the sample. The sub-figures were produced by the CloudBrain-MRS. tNAA = total N-acetylaspartate, tCho = total choline, Glx = glutamate + glutamine, Ins = inositol, GSH = glutathione, tCr = total creatine

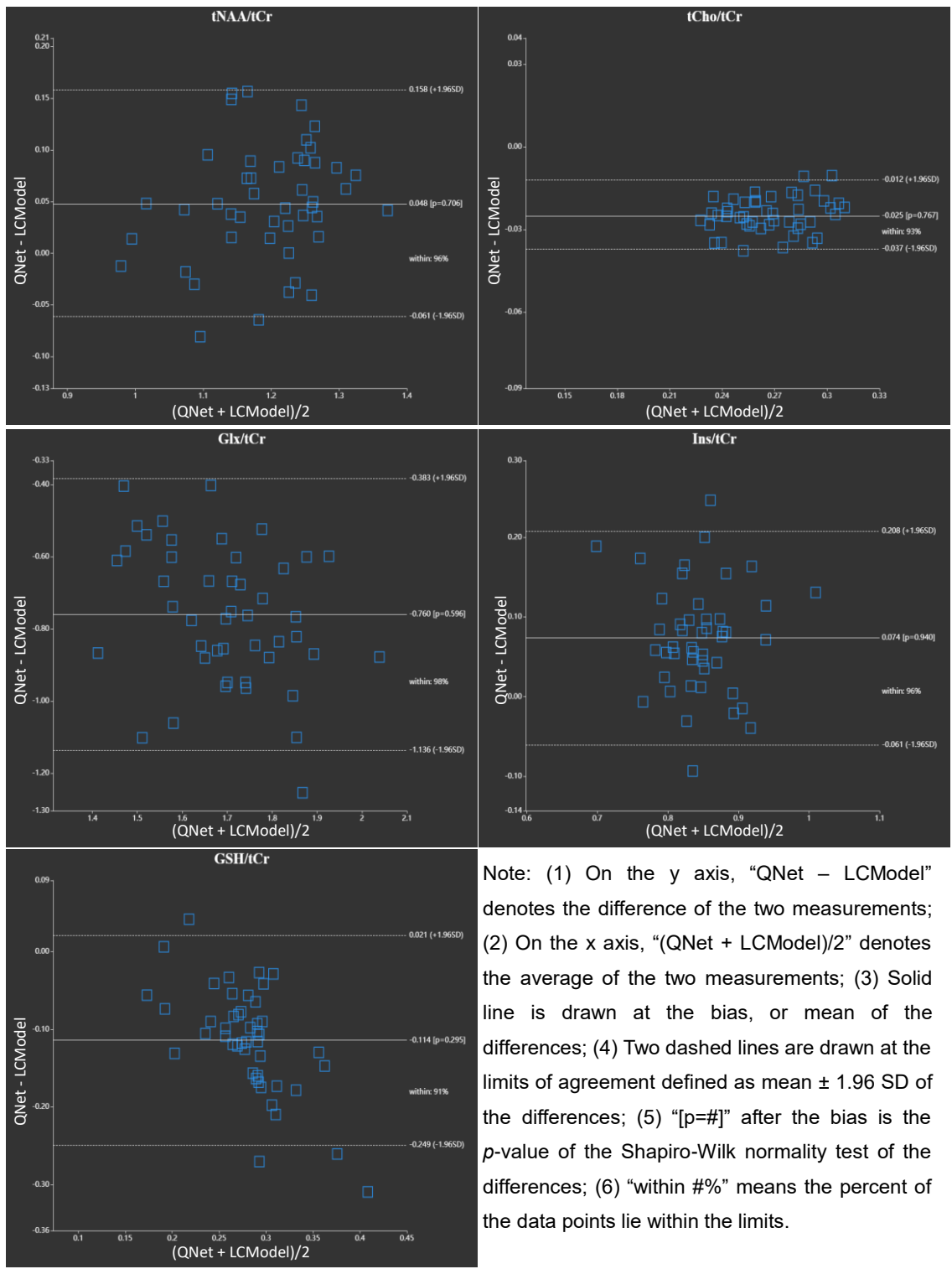

Note: (1) On the y axis, "QNet – LCModel" denotes the difference of the two measurements; (2) On the x axis, "(QNet + LCModel)/2" denotes the average of the two measurements; (3) Solid line is drawn at the bias, or mean of the differences; (4) Two dashed lines are drawn at the limits of agreement defined as mean ± 1.96 SD of the differences; (5) "[p=#]" after the bias is the *p*-value of the Shapiro-Wilk normality test of the differences; (6) "within #%" means the percent of the data points lie within the limits.

**Fig. 4.** Bland-Altman plots for the 46 spectra acquired by scanner Philips Achieva between QNedt and LCModel for quantification of the five metabolite concentrations relative to tCr of tNAA, tCho, Glx, Ins, and GSH. A box indicates a spectrum in the sample. The sub-figures were produced by the CloudBrain-MRS. tNAA = total N-acetylaspartate, tCho = total choline, Glx = glutamate + glutamine, Ins = inositol, GSH = glutathione, tCr = total creatine

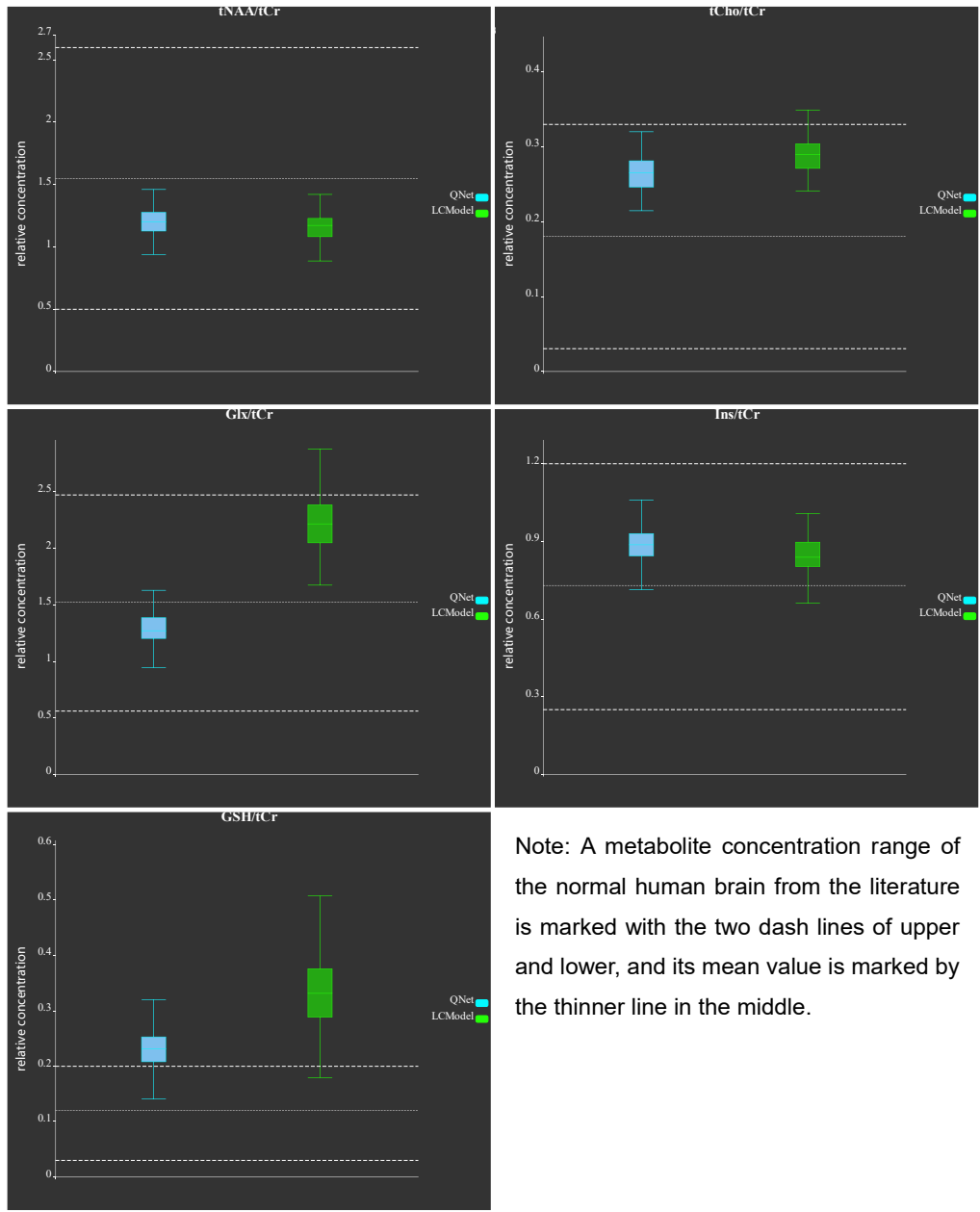

**Fig. 5.** Reasonability analysis illustrated by boxplots of the metabolite concentrations estimated by QNet (blue) and LCModel (green) for the spectra acquired by both scanners. The sub-figures were produced by the CloudBrain-MRS. tNAA = total N-acetylaspartate, tCho = total choline, Glx = glutamate + glutamine, Ins = inositol, GSH = glutathione, tCr = total creatine